\documentclass{elsart}

\usepackage{natbib}

\usepackage{graphicx}

\usepackage{amssymb}

\begin{document}

\begin{frontmatter}



\title{The temperature structure in the core of S\'ersic 159-03}


\author[label1,label2]{J. de Plaa}
\author[label1]{J. S. Kaastra}
\author[label1]{M. M\'endez}
\author[label3]{T. Tamura}
\author[label1]{J. A. M. Bleeker}
\author[label4]{J. R. Peterson}
\author[label5]{F. B. S. Paerels}
\author[label6]{M. Bonamente}
\author[label6]{R. Lieu}

\address[label1]{SRON National Institute for Space Research, Sorbonnelaan 2, 3584 CA Utrecht, The Netherlands}
\address[label2]{Astronomical Institute, Utrecht University, PO Box 80000, 3508 TA Utrecht, The Netherlands}
\address[label3]{Institute of Space and Astronautical Science, JAXA, 3-1-1 Yoshinodai, Sagamihara, Kanagawa 229-8510, Japan}
\address[label4]{KIPAC, Stanford University, PO Box 90450, MS 29, Stanford, CA 94039,USA}
\address[label5]{Department of Astronomy, Columbia University, 550 West 120th Street, New York, NY 10027, USA}
\address[label6]{Department of Physics, University of Alabama, Huntsville, AL 35899, USA}

\begin{abstract}
We present results from a new 120 ks XMM-Newton observation of
the cluster of galaxies S\'ersic 159-03. In this paper we focus on the 
high-resolution X-ray spectra obtained with the Reflection Grating Spectrometer 
(RGS). The spectra allow us to constrain the temperature structure in 
the core of the cluster and determine the emission measure distribution 
as a function of temperature. We also fit the line widths of 
mainly oxygen and iron lines.
\end{abstract}


\end{frontmatter}

\section{Introduction}

X-ray emission from clusters of galaxies is dominated by diffuse
emission from virialized hot plasma which is trapped in the clusters
gravitational potential. Current X-ray missions like XMM-Newton
and Chandra show that the structures in the hot diffuse plasma can be
very complex.

Recent high-resolution X-ray spectra of cool-core clusters by the RGS instrument
aboard XMM-Newton \cite[]{peterson2001,peterson2003} show that the
cores of many clusters lack strong Fe XVII emission lines. This indicates
that the amount of cool gas in the cores is too little to be explained
by classical cooling-flow models \cite[see e.g.][]{fabian1994}. From
the studies of larger samples of cool-core clusters with XMM-Newton
\cite[]{peterson2003,kaastra2004} we learn that the cores are not
isothermal, but probably multi-phase. The implication of the lack of cool gas
is that cool-core clusters need an additional heat source to balance the 
radiative cooling of the gas.

In this paper we discuss the results of a 120 ks XMM-Newton RGS observation
of the cluster of galaxies S\'ersic 159-03. This cluster is relatively nearby
and has a redshift of $z = 0.0564$. We concentrate on the
temperature structure of the core and the widths of the emission lines. 

\section{Spectral models}

The temperature distribution in the complex cores of clusters and the amount of
cool gas, can be determined from its spectrum and does not necessarily have to
be resolved spatially. From the XMM-Newton data of a large sample of clusters
we know that cluster cores can be best fitted with a differential emission measure
(DEM) model \citep[e.g.][]{peterson2003,kaastra2004,deplaa2004}. In this model the
emission measure  ($Y = \int n_e n_H dV$) of a number of thermal components is
distributed as a function of temperature ($T$). This is shown in Eq.~(\ref{eq:dy_dt})
adapted from \citet{kaastra2004}:
\begin{equation}
\frac{dY}{dT} = \left\{ \begin{array}{ll}
cT^{1/\alpha} & \hspace{1.0cm} T < T_{\mathrm{max}}, \\
0 & \hspace{1.0cm} T > T_{\mathrm{max}}. \\
\end{array} \right.
\label{eq:dy_dt}
\end{equation}
This distribution is cut off at some fraction of $T_{\mathrm{max}}$ which is set to 0.1
times $kT_\mathrm{max}$ in this study. The model above is an empirical parametrization 
of the DEM distribution found in the core of many clusters. In this form the limit 
$\alpha \to 0$ yields the isothermal model.

\begin{table*}[tb]
\caption{The model parameters used to produce Fig.~\ref{fig:sim_wdem} and Fig.~\ref{fig:dem_wdem}.}
\begin{center}
\begin{tabular}{|lccccc|}
\hline
{\bf Model} 	& {\bf Linestyle}	& {\bf $kT_{\mathrm{max}}$ }    & {\bf $\alpha$}& {\bf Mean $kT$} & {\bf Cut-off} \\
\hline
DEM		& -~-~-~-~-		&	3.0 keV			& 0.4		& 2.33 keV	& 0.1$kT_\mathrm{max}$ \\ 
DEM		& --~$\cdot$~--~$\cdot$ &	3.0 keV			& 1.0		& 2.02 keV	& 0.1$kT_\mathrm{max}$ \\
DEM		& $\cdots \cdots$	&	3.47 keV		& 1.0		& 2.33 keV	& 0.1$kT_\mathrm{max}$ \\
Cooling-flow	& --------		&	3.0 keV			&		&		& 0.1 keV\\
\hline	
\end{tabular}
\end{center}
\label{tab:legenda}
\end{table*}

\begin{figure}[!t]
\begin{center}
\includegraphics[width=13.5cm,height=10.0cm]{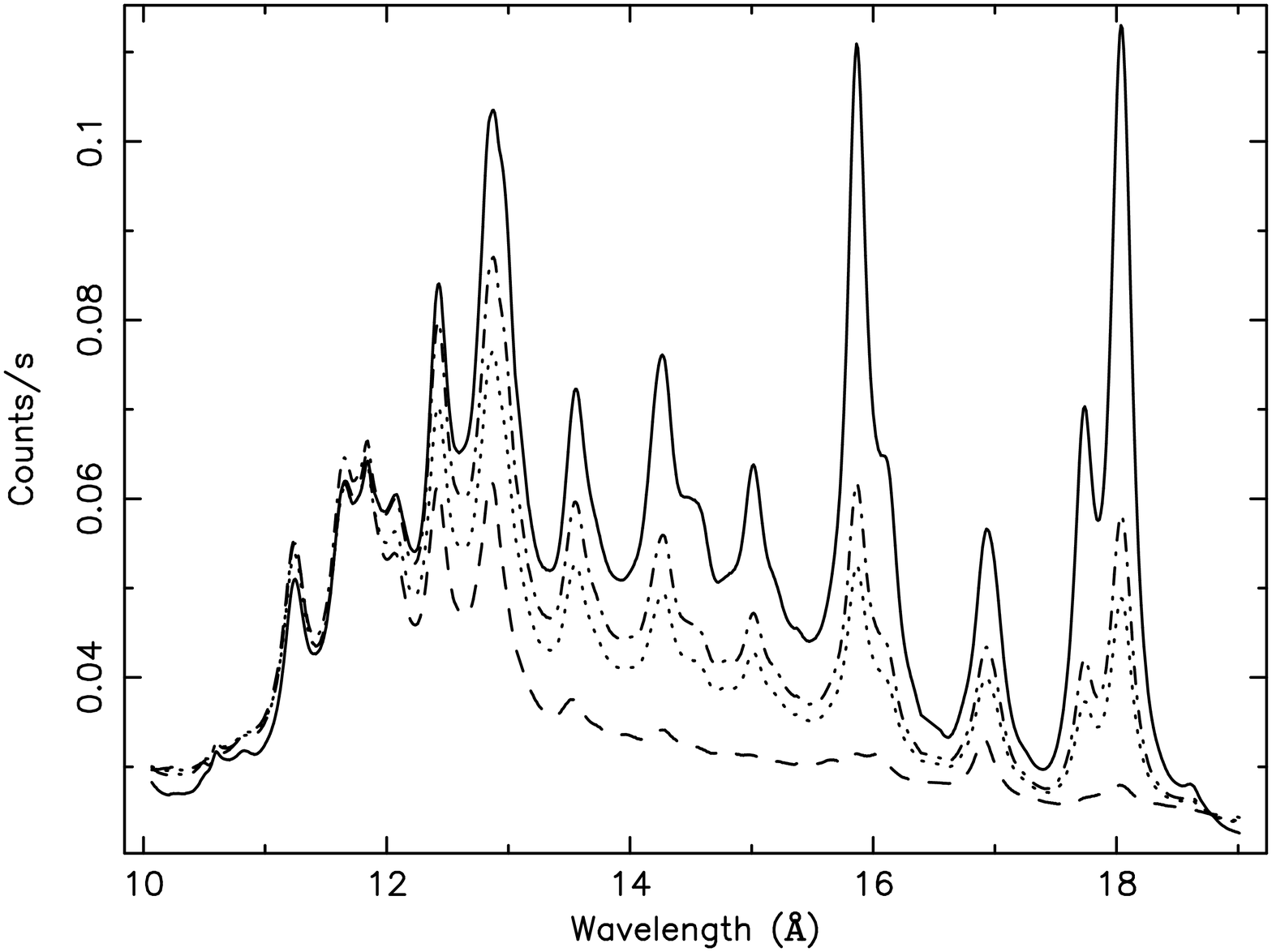}
\end{center}
\caption{Simulated RGS spectra of the Fe-L complex using the model parameters listed in Table~\ref{tab:legenda}.}
\label{fig:sim_wdem}
\vspace{0.3cm}
\begin{center}
\includegraphics[width=13.5cm,height=10.0cm]{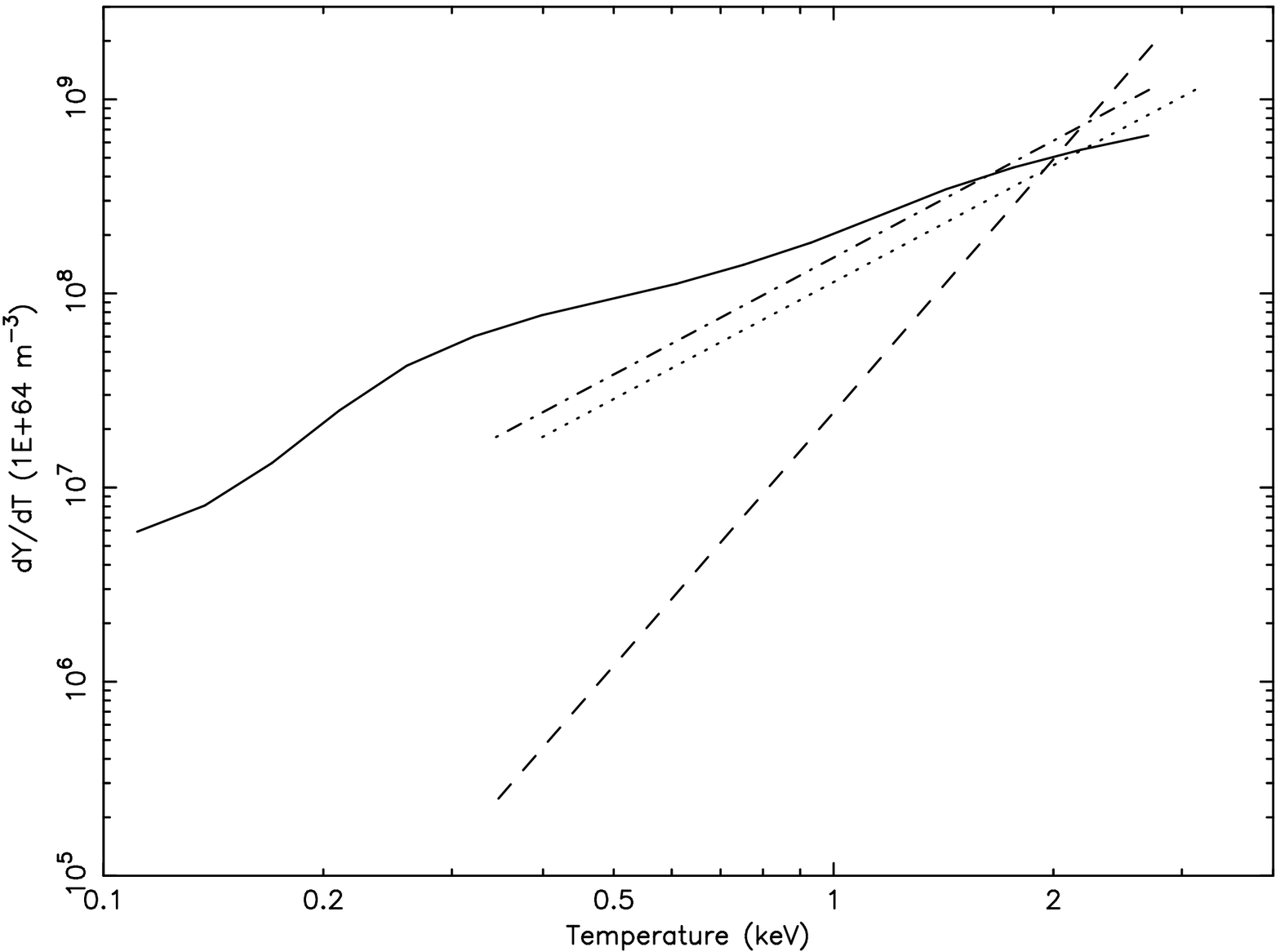}
\end{center}
\caption{DEM distributions for the models listed in Table~\ref{tab:legenda}. 
}
\label{fig:dem_wdem}
\end{figure}

Well resolved spectral lines are important for this model to give an accurate
result, because the strength of each line has a unique dependence on the temperature.
As an example, we show simulated RGS spectra of the Fe-L complex in Fig.~\ref{fig:sim_wdem}. 
These spectra are simulated using four different sets of model parameters listed in 
Table~\ref{tab:legenda}. We assume a cluster spectrum emitted at the redshift of S\'ersic 159-03 
($z$=0.0564). The respective DEM profiles corresponding to these spectra are shown 
in Fig.~\ref{fig:dem_wdem}. From this figure we see that the
model with $\alpha$ = 1.0 has a larger contribution of cold gas than $\alpha$ = 0.4.
In the spectrum (Fig.~\ref{fig:sim_wdem}) this leads to enhanced line emission
above 11 \AA ~from ions like Fe XVII. Thus, by fitting high-resolution spectra with
this model we are able to derive the shape of the temperature distribution. 

The first two models listed in Table~\ref{tab:legenda} show the difference between 
two values for $\alpha$. 
In order to compare the outcome of these models with single-temperature fit results
found in other studies, we also calculate the weighted mean temperature of the DEM distribution.  
The third model in Table~\ref{tab:legenda} is chosen in a way that the mean temperature
of the distribution is the same as the one in the first model. Finally, the fourth model
represents the isobaric cooling-flow model.  

Because the RGS gratings operate without a slit, the resulting spectrum of an extended source is the sum of
all spectra in the (in our case) $1^{\prime} \times \sim 12^{\prime}$ field of view, convolved with the PSF
\cite[for a complete discussion about grating responses see][]{davis2001}.
Extended line-emission appears to be broadened depending on the spatial extent of the source along the dispersion direction.
In order to describe the data properly, the spectral fits need to account for this effect. In practice, this is accomplished by
convolving the spectral models with the surface brightness profile of the source along the dispersion direction. 
For that purpose we extract the cluster intensity profile from MOS1 along the dispersion direction of RGS,
which we convolve with the RGS response during spectral fitting. This procedure is 
described in \cite{tamura2004} and was also applied in studies like, for example, \cite{kaastra2001} and \cite{deplaa2004}.
Because the radial profile of an ion can be different from the wide-band surface brightness profile, this method is not ideal. 
In order to account for this, we let the scale of the width and the position of the profile free in the fit to match the 
profiles of the main emission lines.

\begin{figure*}[t]
\begin{center}
\includegraphics[width=13.5cm,height=10.0cm]{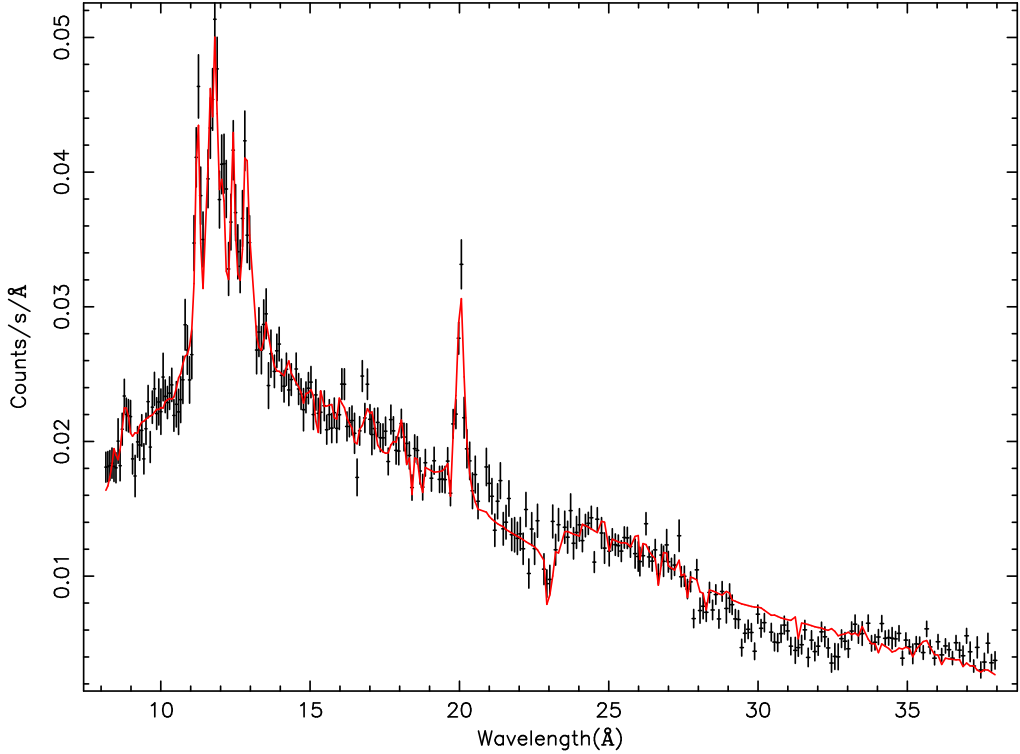}
\end{center}
\caption{RGS spectrum of S\'ersic 159-03 fitted with a DEM model with a variable width
for the oxygen and iron lines. The wavelength plotted here is the observed (redshifted) wavelength.
From the left to the right we see emission lines from Mg (9 \AA), Fe-L and Ne (11--14 \AA) and O (20 \AA).}
\label{fig:rgs}
\end{figure*}

\section{Results}

\begin{table*}[t]
\caption{Best fit results from a DEM model fit to the spectrum extracted from a 4 arcmin wide
extraction region. The line width relative to the FWHM of the cluster line profile  
is also given. This cluster line profile derived from the clusters surface brightness profile 
has a FWHM of 0.14 \AA.
}
\begin{center}
\begin{tabular}{|lr|lr|}
\hline
{\bf Parameter}         & {\bf Value}           & {\bf Parameter}       & {\bf Value} \\
\hline
$kT_{\mathrm{max}}$ (keV) & 4.08 $\pm$ 0.16     & cut off               &  0.05 $\pm$ 0.04 \\
$\alpha$                & 0.43 $\pm$ 0.03       & mean $kT$             &  3.14 $\pm$ 0.13 \\
\hline
Line width (C,N,O)      & 1.5 $\pm$ 0.3         & Line width (Ne,Mg,Fe)   & 0.57 $\pm$ 0.07 \\
\hline
\end{tabular}
\end{center}
\label{tab:results}
\end{table*}

In Table~\ref{tab:results} we show the results of a fit to the RGS spectrum of S\'ersic 159-03 
shown in Fig.~\ref{fig:rgs}. We extract this spectrum from a 4 arcmin wide region 
in the cross-dispersion direction of the RGS.  The spectrum shows line
emission from Fe-L, Ne, Mg and O. In the region between $\sim$ 14--19 \AA~no 
strong features from Fe XVII lines can be observed and therefore the fitted value of 
$\alpha$ is moderate: 0.43 $\pm$ 0.03. Together with the $kT_{\mathrm{max}}$ from the
fit we find a mean temperature of the core of 3.14 $\pm$ 0.13 keV. This value is not 
so different from the value of 2.80 $\pm$ 0.08 keV obtained when fitting a single-temperature 
model to the data.

\begin{figure}
\includegraphics[width=1.0\textwidth]{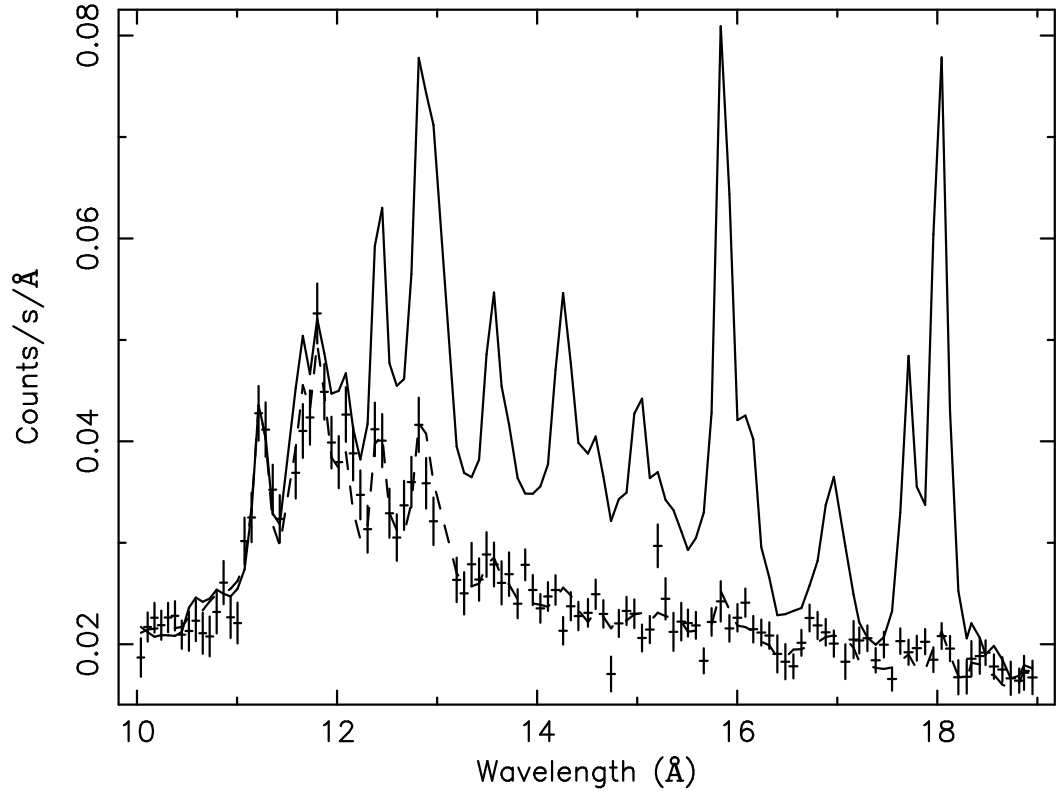}
\caption{RGS combined spectrum of the 10--19 \AA~wavelength range. The dashed line represents
the best-fit DEM model and the continuous line shows a typical cooling-flow model.}
\label{fig:coolflow}
\end{figure}

When we look into the 10--19 \AA~wavelength range and compare the best fit DEM model
with the classical cooling-flow model, we obtain a spectrum like in Fig.~\ref{fig:coolflow}.
The lower and upper temperature limits of the cooling-flow model (continuous line) have 
been set to 0.1 and 4.0 keV respectively. The dashed-line represents the best fit DEM model 
and is the same as the model presented in Fig.~\ref{fig:rgs}. From this plot it is clear
that the iron lines associated with the cold gas, which are predicted by the cooling-flow 
model, are not observed.

In Fig.~\ref{fig:rgs} we can see by eye that the oxygen line near 20~\AA~is broader
than the individual iron lines in the Fe-L complex at around 11--13 \AA. A fit to the width of the lines 
confirms this view and shows that the line width of oxygen is nearly 3 times larger
than the width of the iron lines. The lines of the other fitted elements (carbon, nitrogen, 
neon and magnesium) do not have a large signal-to-noise, therefore we couple their widths to
either iron or oxygen. The line widths of carbon and nitrogen are coupled to oxygen, while neon and magnesium 
are coupled to iron. In Table~\ref{tab:results} we show the widths of these lines with respect to the 
broad band cluster profile.

\section{Conclusion}

We performed a spectral analysis of the cool-core cluster S\'ersic 159-03 using the
RGS instrument aboard XMM-Newton. The high-resolution spectra of the core confirm
the lack of cool gas in the core of the cluster in agreement with the results found by 
\cite{peterson2001} and \cite{peterson2003} in this and other clusters.
From the line width of O and Fe we can derive the spatial distribution of the elements.
Fe has a line width of 0.57 $\pm$ 0.07 with respect to the profile of the continuum emission
derived from EPIC MOS, while the O line is much broader with a value of 1.5 $\pm$ 0.3. This
means that the Fe distribution in the cluster is much more centrally peaked than the oxygen
distribution. This is consistent with earlier measurements by e.g. \cite{tamura2004}.

\bibliographystyle{elsart-harv}
\bibliography{asr}

\end{document}